# Janus Bound States in the Continuum with Asymmetric Topological Charges

Meng Kang[1], Meng Xiao[2,3,†], and C. T. Chan[1,‡]

[1]*Department of Physics, The Hong Kong University of Science and Technology, Hong Kong, China*

[2]*Key Laboratory of Artificial Micro- and Nano-structures of Ministry of Education and School of Physics and Technology, Wuhan University, Wuhan 430072, China*

[3]*Wuhan Institute of Quantum Technology, Wuhan 430206, China*

We propose a novel topological defect called Janus bound states in the continuum (BICs), featuring asymmetric topological charges in upward and downward radiation channels. Our approach involves a photonic crystal slab (PCS) that initially exhibits both out-of-plane and in-plane mirror symmetry, and this PCS possesses one BIC at the $\Gamma$ point and two BICs off the $\Gamma$ point. By introducing certain perturbations that break the out-of-plane mirror symmetry, the two off-$\Gamma$ BICs decompose into four circularly polarized states (C points) with identical topological charges (each with half the topological charge of the original BIC) while the at-$\Gamma$ BIC is preserved. Then, we selectively manipulate the four C points associated with the downward radiation channel to converge at the at-$\Gamma$ BIC, forming a Janus BIC with distinct topological charges for upward and downward radiation. By further introducing in-plane mirror symmetry perturbation, we can bring two of the C points with the same handedness and identical topological charges for upward radiation to merge into the Janus BIC. This process results in a Janus chiral BIC which exhibits large intrinsic chirality and an infinite Q factor. Janus BICs can induce distinct Pancharatnam–Berry phase singularities in momentum space for different incident channels, providing a new approach to control orbital angular momentum. Janus chiral BICs hold promise in enhancing direction-dependent and spin-dependent asymmetric light-matter interaction, opening new pathways for improving chirality-dependent operation for on-chip devices.

Introduction.—Topological defects with fascinating properties have attracted significant attention in photonics recently, including bound states in the continuum (BICs) [1-11], unidirectional guided resonances (UGRs) [12-14] and circularly polarized states (C points) [15-18]. These defects manifest as polarization vortices characterized by nonzero topological charges in momentum space. BICs with integer topological charges [FIG. 1(a)] exhibit strong light confinement with an infinite quality (Q) factor and the ability to induce sharp Fano resonances [10,11]. On the other hand, UGRs with integer topological charges on either the up or down side showcase unidirectional radiation [12], while C points with half-integer topological charges possess large chirality [15,16,18]. By exploiting these topological defects to enhance light-matter interaction, a wide range of applications have been developed [11], including lasing [19-22], nonlinear optical effects [23,24], sensing [25,26], exciton–polaritons [8,9,27], chiral light sources[28], grating couplers [29] and vortex beam generation [30,31]. New topological defects, such as Janus BICs with distinct topological charges on each side of a slab [FIG. 1(b)], will exhibit new properties and offer new applications. For instance, Janus BICs can exist even in the absence of any mirror symmetries, which can overcome the challenge of realizing chiral BICs that exhibit large intrinsic chirality while retaining an infinite Q factor. In addition, the unique topological feature of Janus BICs provides a novel way to manipulate phase singularities, which can be useful in various fields, including optical communications and quantum information processing [32]. The presence of out-of-plane asymmetric polarization vortices makes Janus BICs promising for applications in nonlinearity-induced nonreciprocal devices [33,34]. Nevertheless, the existence of Janus BICs remains elusive. Creating a Janus BIC necessitates the breaking of out-of-plane mirror symmetry to generate asymmetric charges, which makes the adjustable independent parameters in the system insufficient for eliminating radiation. While such symmetry breaking typically results in the destruction of BICs [1,12], we demonstrate here that large intrinsic chirality and infinite Q can co-exist at normal incidence.

Enhancing chirality has long been a sought-after goal to advance applications reliant on chiral light-matter interaction [35]. Although C points exhibit large optical chirality, their emergence typically deviates from the normal direction [12-16], resulting in large *extrinsic* chirality [17,36]. Consequently, applications requiring *intrinsic* chirality, including chiral light sources and chiral photodetectors, have faced limitations. To address this challenge, chiral quasi-BICs with large intrinsic chirality and high Q factors have been proposed [18,37-39]. However, the formation of chiral quasi-BICs necessarily breaks the BIC condition, imposing an upper limit on the achievable Q factor. On the other hand, the infinite Q factor exhibited by true BICs means the strongest possible light confinement. Therefore, the realization of chiral BICs, characterized by large intrinsic chirality and infinite Q factors, is crucial for boosting chiral light-matter interaction. However, currently, there is no available approach to achieve chiral BICs. Unlike conventional BICs, Janus BICs possess the unique ability to withstand the breaking of necessary mirror symmetries as required by chiral BICs. Thus, Janus BICs enable a possible practical implementation of chiral BICs.

In this Letter, we introduce the concept of Janus BICs, which is a new type of topological defect featuring asymmetric topological charges in upward and downward radiation. By harnessing the unique topological property of Janus BICs, we achieve chiral BICs and gain additional freedom in controlling phase singularities. Specifically, Janus BICs are produced by introducing up-down mirror symmetry ($\sigma_z$) breaking to merge multiple C points with the same topological charges into a symmetry-protected BIC on only the downward side. Consequently, the topological charge of the BIC on the downward side is modified, while the upward side remains unaffected, forming a Janus BIC. To achieve chiral BICs, we further introduce in-plane mirror symmetry breaking to tune two C points with the same topological charge and identical handedness on the upward side to the Janus BIC. This process ends in a Janus chiral BIC with large intrinsic chirality and an infinite Q factor. We also demonstrate the capability of Janus BICs in inducing distinct Pancharatnam–Berry phase singularities for direction-dependent reflection and transmission.

Janus BICs.—Our construction of Janus BICs starts from a photonic slab that supports both symmetry-protected BICs [40-42] at the $\Gamma$ point and accidental BICs [1] at off-$\Gamma$ points. Such a photonic slab exhibits both in-plane two-fold rotational symmetry $C_2^z$ and out-of-plane mirror symmetry $\sigma_z$. When $\sigma_z$ is broken, off-$\Gamma$ BICs split into pairwise C points, whereas at-$\Gamma$ BICs are protected by $C_2^z$ and remain undamaged, resulting in their topological charge being unchanged. Since there is no $\sigma_z$ relating the C points for upward and downward radiation, C points in either radiation direction can be tuned independently with different evolution trajectories. Subsequently, multiple C points on either the upward or downward radiation can be tuned to merge into the symmetry-protected BIC by varying a proper structural parameter. When the total topological charge of those multiple C points does not cancel out at the merging point, a Janus BIC with asymmetric topological charges is achieved.

Following the scheme above, we present a numerical demonstration of a Janus BIC realized using a bilayer photonic crystal slab (PCS) possessing $C_{2v}$ point group symmetry. The PCS comprises a freestanding bilayer $Si_3N_4$ slab (refractive index $n$ =2.02) etched with elliptic cylindrical holes arranged in a square lattice [Fig. 2(a)], which is realizable through nanofabrication technologies such as electron-beam lithography and reactive-ion etching [19,26,29,43]. Two PCS layers have the same thickness (*t*/2) and elliptic holes with matching aspect ratios $\beta$. The minor axes of the elliptic holes on the top and bottom layers are $d+\Delta d$ and $d-\Delta d$, respectively. Thus, $\sigma_z$ can be broken by choosing $\Delta d \neq 0$. When $\Delta d = 0$, the bilayer PCS supports a symmetry-protected BIC at the $\Gamma$ point and off-$\Gamma$ BICs along the $\Gamma$X direction on the $TM_1$ band [Fig. 2(b, c)]. Owing to the anisotropy of elliptic cylindrical holes, well-selected structural parameters can prevent the co-occurrence of off-$\Gamma$ BICs in the $\Gamma X'$ directions [Fig. 2(c) and Supplemental Materials [44]]. When $\Delta d \neq 0$, the broken $\sigma_z$ breaks the off-$\Gamma$ BICs into pairs of C points with opposite handedness but the same topological charges (each is half of the charge of the original off-$\Gamma$ BIC). Correspondingly, the Q factor decreases from infinity to a finite number [Fig. 2(c)]. Meanwhile, the at-$\Gamma$ BIC is preserved during the variation of $\Delta d$.

To reveal the formation of a Janus BIC, we investigate the evolution of topological charges of far-field polarizations for both upward and downward radiation. BICs and C points are identified as topological defects of the polarization vortices in momentum space [2,15]. As shown in Fig. 2(d), when $\Delta d \neq 0$, the off-$\Gamma$ BICs split into pairwise C points. Then with an increase in $\Delta d$, the four C points for downward radiation gradually approach each other, and eventually merge into the at-$\Gamma$ BIC at $\Delta d = 10$ nm. Meanwhile, the four C points for upward radiation remain separate. The far-field polarization distributions in momentum space are presented in Fig. 2(e). When $\Delta d = 0$ nm, the at-$\Gamma$ BIC has a topological charge $q = +1$, whereas the two off-$\Gamma$ BICs connected by $C_2^z$ have topological charges $q = -1$. The polarization fields exhibiting an extremely weak chirality are identical for upward and downward radiation as ensured by $\sigma_z$. As $\Delta d$ increases, the at-$\Gamma$ BIC is preserved, while the two off-$\Gamma$ BICs split into two pairs of C points with the same topological charge $q = -1/2$ and opposite handedness as indicated by different colors. Since $\sigma_z$ is broken, C points for upward and downward radiation have distinct distributions. At $\Delta d = 10$ nm, four C points with the same topological charge $-1/2$ merge into the at-$\Gamma$ BIC for downward radiation, while all the C points become more separated for upward radiation. Consequently, we obtain a topologically stable Janus BIC which exhibits topological charges $q^{\mathrm{u}} = +1$ and $q^{\mathrm{d}} = -1$ for the upper and lower sides, respectively.

We further demonstrate the preservation of Janus BICs in the absence of all mirror symmetries. One of the most convenient methods to break the residual in-plane mirror symmetry is to rotate the elliptic cylindrical holes. When the upper-layer holes are rotated [Fig. 3(a)], C points with opposite handedness are no longer constrained by mirror symmetry. Due to the remaining $C_2^z$, C points with the same topological charge and handedness move in pairs in momentum space as the rotated angle $\alpha^{\mathrm{u}}$ varies [Fig. 3(b)]. Additionally, since the at-$\Gamma$ state exhibits a different $C_2^z$ eigenvalue compared to the free propagating states, it cannot radiate either upward or downward, i.e., a BIC is preserved at $\Gamma$ independent of $\alpha^{\mathrm{u}}$ [e.g., Fig. 3(c)]. When $\alpha^{\mathrm{u}} = 4.5^\circ$, the two right-handed C points are tuned to $\Gamma$ for upward radiation, while the two left-handed C points for upward radiation move away from $\Gamma$ as $\alpha^{\mathrm{u}}$ increases. Meanwhile, the four C points for downward radiation remain pinned at $\Gamma$. Consequently, the topological charges of the Janus BIC become $q^{\mathrm{u}} = 0$ and $q^{\mathrm{d}} = -1$. With further increase of $\alpha^{\mathrm{u}}$, the two right-handed C points for downward radiation start to move away from $\Gamma$ at $\alpha^{\mathrm{u}} = 5^\circ$ [the blue solid lines in Fig. 3(b)]. Eventually, the BIC at $\Gamma$ exhibits zero topological charges for both upward and downward radiation, $q^{\mathrm{u}} = q^{\mathrm{d}} = 0$. Similarly, another way to break in-plane mirror symmetry is to rotate holes on both layers by the same angle. In this scenario, one pair of C points with the same handedness is tuned to merge into a Janus BIC for upward radiation (Supplemental Materials [44]).

Intrinsic chirality.—We proceed to show the process above enables the formation of chiral BICs with large intrinsic chirality and an infinite Q factor. Intrinsic chirality is realized by tuning C points with large optical chirality to the Γ point, which requires the breaking of all mirror symmetries [18]. Stokes parameter $S_3$ of the far-field radiation can be utilized to reveal optical chirality [18,46]. As previously discussed, rotating elliptic cylindrical holes introduce in-plane mirror symmetry breaking, enabling C points with the same handedness to be tuned to Γ. Figure 3(d) shows the polarization distribution at $\alpha^{\mathrm{u}} = 5^\circ$. At this critical point, the two right-handed C points for downward radiation start to move away from Γ, while the two right-handed C points for upward radiation remain merging at Γ. Consequently, a chiral BIC with large intrinsic chirality in both radiation directions is produced at $\alpha^{\mathrm{u}} = 5^\circ$. The $S_3$ distribution approaches $-1$ for both upward and downward radiation near the chiral BIC [Fig. 3(e)]. This can be seen more clearly by tracking the minimum of $S_3$ as a function of $k_x$ which reaches -1 at $k_x = 0$ [Supplemental Materials [44]]. Notable, since light is perfectly confined in the PCS with an infinite Q at Γ [Fig. 3(c) and Supplemental Materials [44]], $S_3$ is ill-defined at exactly the chiral BIC point. This explains the abrupt change in $S_3$ near Γ. To enable $|S_3| = 1$ in the normal direction, further symmetry perturbations can be introduced to reduce Q and achieve chiral quasi-BICs [Supplemental Materials [44]]. At $\alpha^{\mathrm{u}} = 4.5^{\circ}$ where the two right-handed C points for upward radiation merge at Γ, another chiral BIC is formed, exhibiting large intrinsic chirality for upward radiation.

The magical features of Janus BICs and chiral BICs can lead to unique possibilities for applications. A single Janus BIC can offer multiple functionalities in controlling phase singularity in momentum space; a chiral BIC exhibits an infinite Q factor with a large intrinsic chirality which is critical in chiral emission, sensing and enantiomer separation.

Control phase singularities.—The manipulation of phase singularities in generating optical vortices holds significant importance in various fields, including optical communications and quantum information processing [32]. Polarization vortices can create a phase singularity in momentum space through spin-orbit interaction, leading to promising applications in vortex beam generation [30] and beam shift [47,48]. This approach offers the advantage of circumventing the optical misalignment problem typically associated with inhomogeneous non-periodic nanostructures since structures that are periodic in real space do not have an optical axis. Specifically, polarization vortices induce a geometric phase known as the Pancharatnam-Berry phase [49,50] through polarization conversion, which varies as polarization directions (upside $\theta^{\mathrm{u}}$ and downside $\theta^{\mathrm{d}}$ for the long-axis direction of elliptical polarization) change within momentum space (see detailed derivation in Supplementary Material [44]). Notably, since a Janus BIC exhibits different topological charges for upward and downward radiation, it can induce different polarization direction distributions $\theta^{\mathrm{u}}$ and $\theta^{\mathrm{d}}$. The corresponding polarization conversion from right-handed circularly polarized light (RCP, indicated by "–") to left-handed circularly polarized light (LCP, indicated by "+") under near-normal incidence exhibits a nontrivial phase winding as

$$R_{+-}^{\mathrm{f}} \sim \exp(-i2\theta^{\mathrm{d}}), \tag{1}$$

$$R_{+-}^{\mathrm{b}} \sim \exp(-i2\theta^{\mathrm{u}}), \tag{2}$$

$$T_{+-}^{\mathrm{f}} = T_{+-}^{\mathrm{b}} \sim \exp[-i(\theta^{\mathrm{u}} + \theta^{\mathrm{d}})] \tag{3}$$

Here superscripts 'f' and 'b' denote forward (+*z*) and backward (-*z*) propagation, respectively. The efficiency of polarization conversion, determined by the Stokes parameters of polarization vortices, can be tuned and enhanced by using structural parameters and features found through optimization techniques [Supplementary Material [44]]. Consequently, a Janus BIC can give rise to three spiral phases in momentum space through polarization conversion, with each phase singularity possessing a distinct topological charge of $l = -2q^{d}$, $l = -2q^{u}$, and $l = -(q^{u} + q^{d})$. In the case of polarization conversion from LCP to RCP, the topological charge becomes $-l$. As illustrated in Fig. 4(a), the topological charges of the light field phase singularity depend on reflection, transmission, and the incidence direction. When the topological charge of a Janus BIC changes, the topological charges of the phase singularities undergo corresponding variations according to Eqs. (1-3).

Large intrinsic chirality and infinite Q factor.—Circular dichroism (CD) is a useful indicator to calibrate the chirality of a state. Here CD is defined as $\mathrm{CD}=\frac{(|R_{++}|^2+|R_{-+}|^2)-(|R_{+-}|^2+|R_{--}|^2)}{(|R_{++}|^2+|R_{-+}|^2)+(|R_{+-}|^2+|R_{--}|^2)}$. The reflection matrix $R=[R_{++},R_{+-};R_{-+},R_{--}]$ under forward propagating incidence is (see details in Supplemental Materials [44])

$$R^{\mathrm{f}}=\begin{pmatrix} r+\frac{d_+^d d_-^d}{i(\omega_0-\omega)+\gamma_0} & \frac{\left(d_+^d\right)^2}{i(\omega_0-\omega)+\gamma_0} \\ \frac{\left(d_-^d\right)^2}{i(\omega_0-\omega)+\gamma_0} & r+\frac{d_+^d d_-^d}{i(\omega_0-\omega)+\gamma_0} \end{pmatrix}, \quad (4)$$

where $d_\pm^{\mathrm{d}}$ is the coupling coefficient of downward radiation with LCP/RCP incident wave, *r* is the direct reflection, and $\omega_0$ and $\gamma_0$ are resonance frequency and decay rate, respectively. For backward propagating incidence, $d_\pm^{\mathrm{d}}$ is replaced by $d_\pm^{\mathrm{u}}$ describing upward radiation. The difference between $d_+^{\mathrm{d}}$ and $d_-^{\mathrm{d}}$ causes asymmetric polarization conversion, leading to complete polarization conversion and asymmetric transmission [51,52] (Supplemental Materials [44]) near a chiral BIC. The at-Γ state is a BIC, resulting in reflection spectra without any resonance behavior. To utilize the intrinsic chirality of chiral BICs, perturbation needs to be introduced to enable coupling with circularly polarized radiation. For example, near the chiral BIC, one can easily locate a wavevector with large intrinsic chirality and a high Q factor. Fig. 4(b) shows the simulated CD distribution near the chiral BIC, which approximately reaches −1 in the first and third quarters. By selecting the wavevector indicated in Fig. 4(b), Fig. 4(c) shows the reflection spectra. Complete polarization conversion from RCP to LCP emerges in the reflection [the blue line in Fig. 4(c)], while other reflection coefficients are almost zero, indicating large chirality with a CD of −0.99. The reflection exhibits a sharp Fano resonance with $\mathrm{Q}=1.2\times10^5$.

Summary.—In summary, we have introduced a novel type of topological singularity in momentum space called Janus BICs with asymmetric topological charges. By arranging C points with the same handedness to merge into a Janus BIC, we can achieve chiral BICs with infinite Q factors and large intrinsic chirality. In addition to the mechanism discussed above, Janus BICs can also be constructed by leveraging the interaction between topologically distinct BICs within two separate PCSs. For example, by tuning the separation, a Janus BIC with topological charges of +1 for upward radiation and -2 for downward radiation is achieved in a bilayer PCS with a triangular lattice (Supplemental Materials [44]). Janus BICs offer a new approach to controlling phase singularities in momentum space. Our findings offer a new platform to enhance both direction-dependent and spin-dependent asymmetric light-matter interaction. For instance, the utilization of Janus chiral BICs can enhance chiral light-matter interaction, which is typically weak, leading to advanced performance in areas such as chiral light sources, chiral sensing, and chiral photodetectors. In addition, Janus BICs hold promising applications in vortex beam generation, thereby enhancing the capacity of optical communications and quantum information processing. Owing to out-of-plane asymmetric polarization vortices, Janus BICs offer advantages in nonlinearity-induced nonreciprocal devices, such as nonreciprocal polarization vortices demonstrated in Supplemental Materials [44].

Acknowledgment.—M.K. and C.T.C acknowledge support from Research Grants Council Hong Kong through CRS_HKUST601/23 and AoE/P-502/20. M.X. acknowledges support from the National Natural Science Foundation of China (grant no. 12321161645 and grant no. 12274332).

---

†phmxiao@whu.edu.cn

‡phchan@ust.hk

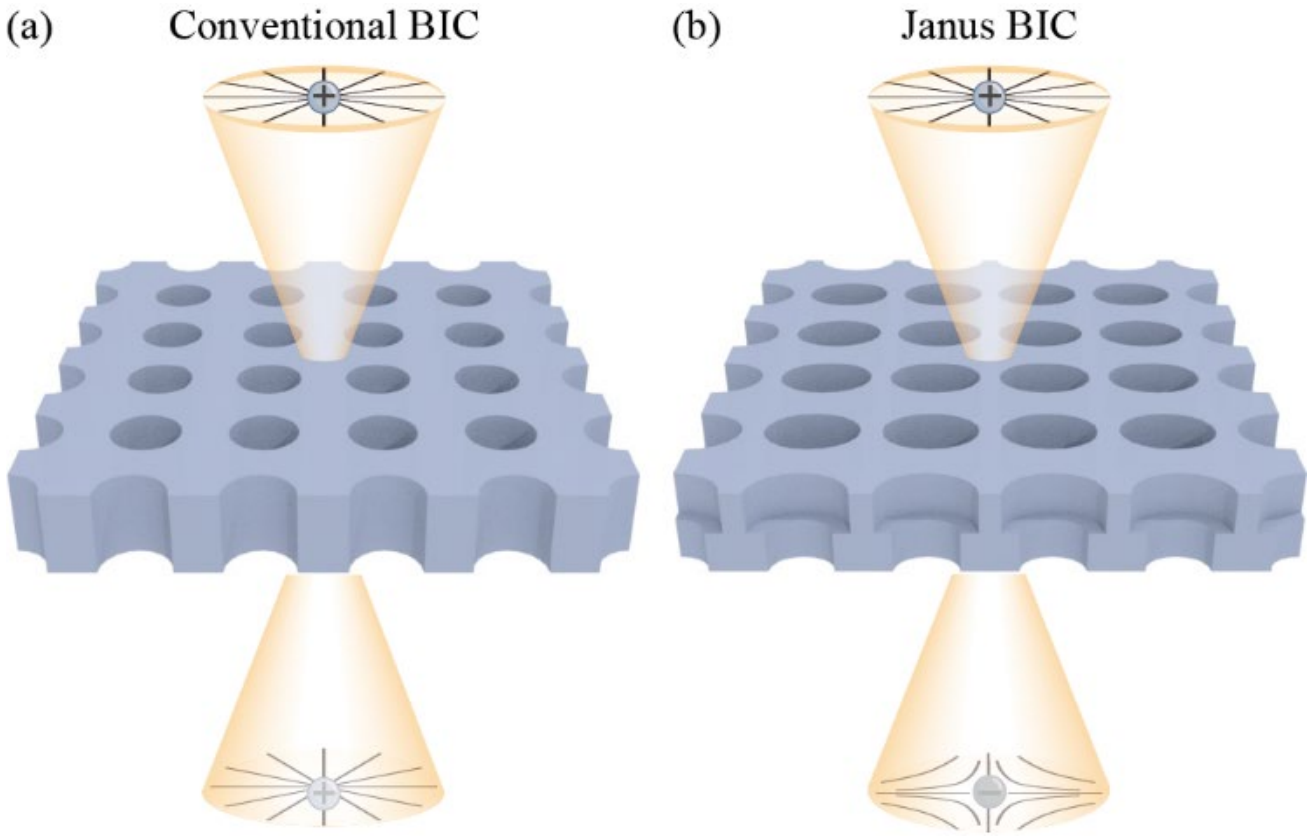

FIG. 1. Sketch illustrating (a) a conventional BIC with symmetric topological charges and (b) a Janus BIC with asymmetric topological charges for upward and downward radiation.

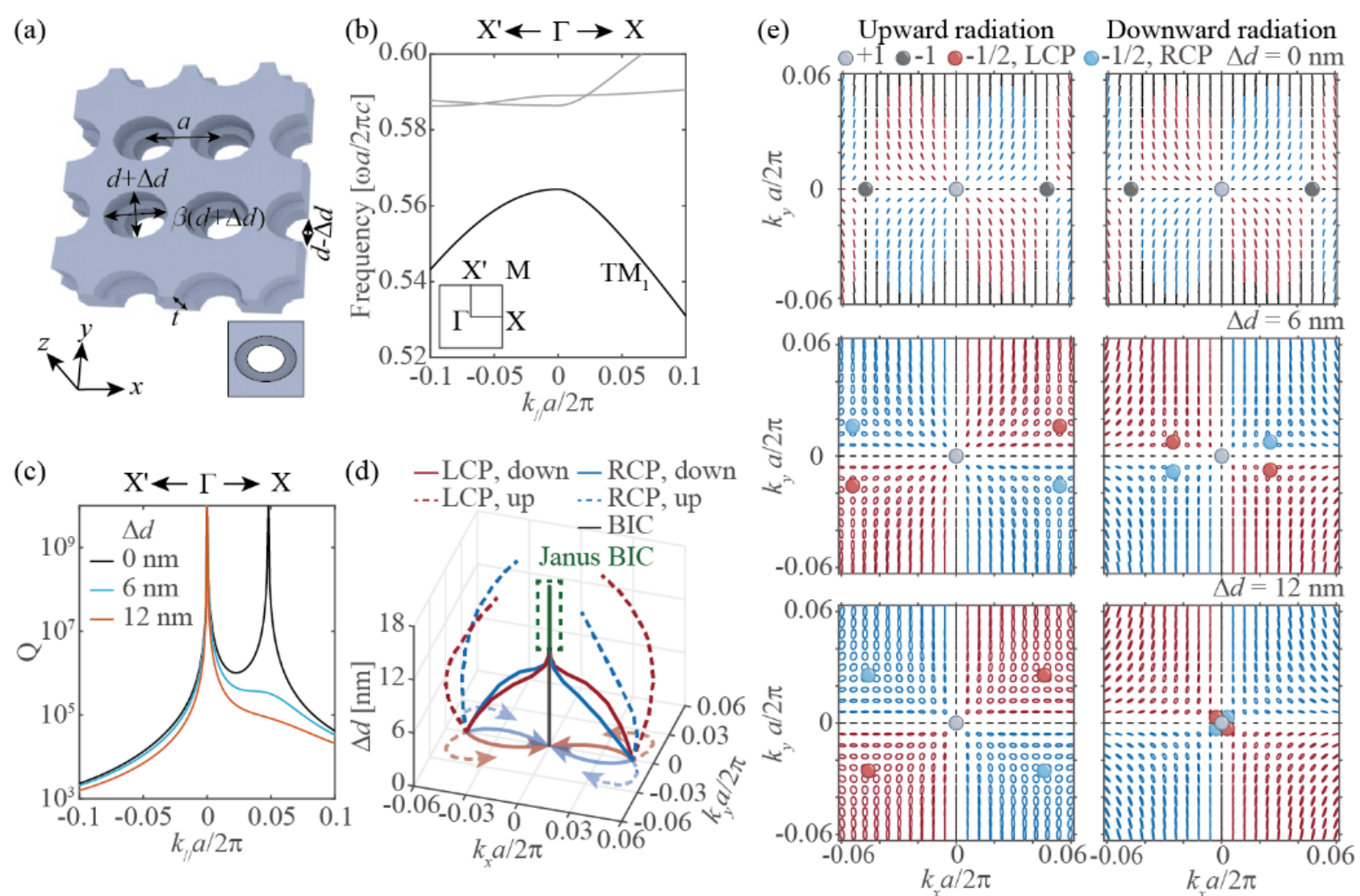

FIG. 2. (a) Schematic of a bilayer PCS designed to realize a Janus BIC. The geometric parameters are: $a$ = 336 nm, $d$ = 0.6$a$, $\beta$ = 1.35, $t$ =480 nm. (b) Calculated TM-like band structure for the case of $\Delta d$ = 0 nm. $\boldsymbol{k}_{//} = (k_x, k_y)$. (c) Calculated Q factors for three different values of $\Delta d$. (d) Evolution of the BIC (the black line) and C points (the red and blue lines) for the upward (dashed lines) and downward (solid lines) radiations. A Janus BIC (the green line) started to emerge at $\Delta d = 10$ nm (highlighted by the green box). The projected arrows show the trajectories of the C points as we increase $\Delta d$. (e) Calculated polarization vortices for three different $\Delta d$. The vortex pattern at $\Delta d = 12$ nm clearly demonstrates a Janus BIC where the topological charge for upward radiation and downward radiation are +1 and -1, respectively.

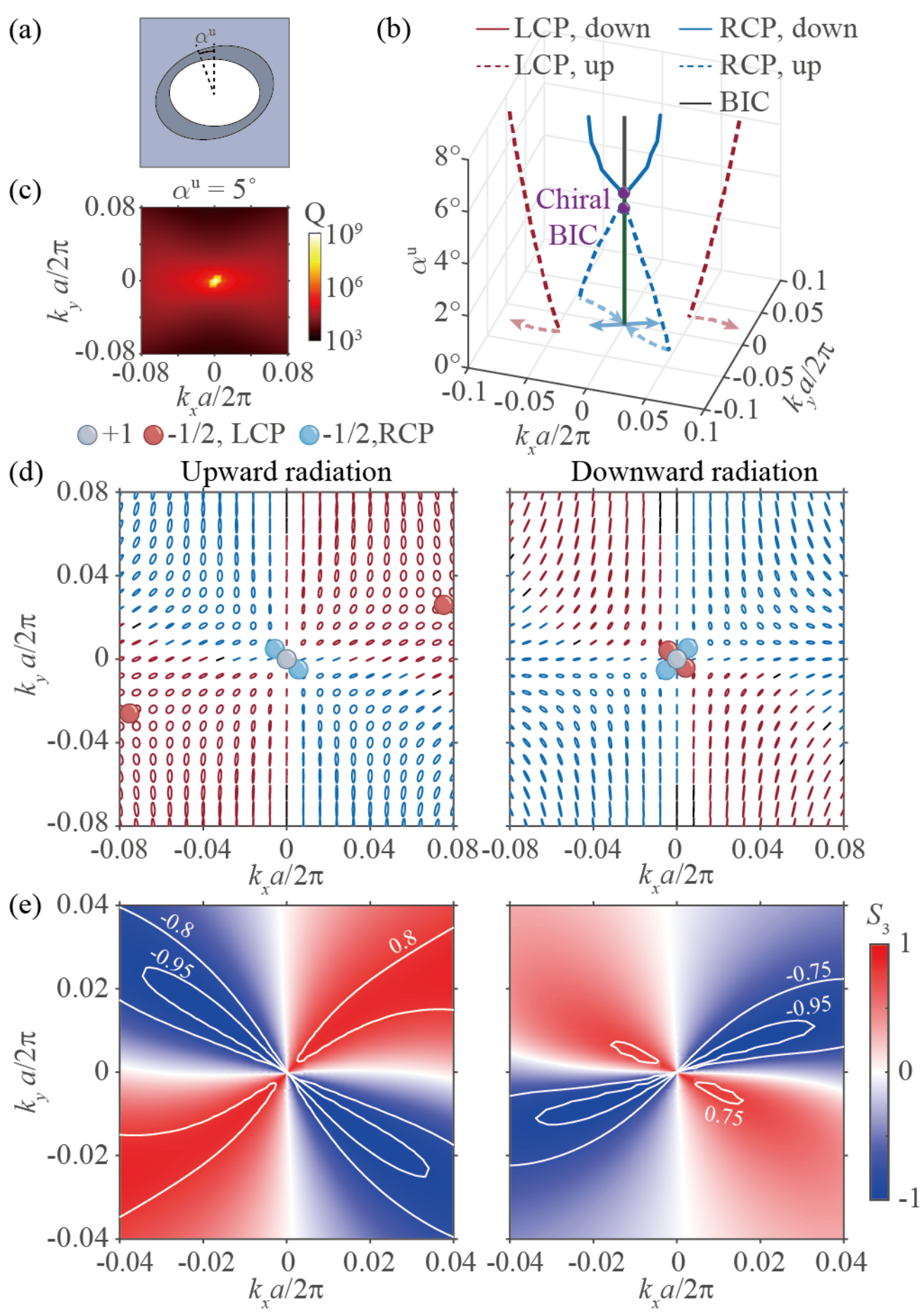

FIG. 3. (a) Diagram illustrating a unit cell with upper-layer holes rotated. (b) Evolution of Janus BICs (the green line) and C points (the red and blue lines) for the upward (dashed lines) and downward (solid lines) radiations. Chiral BICs (the magenta dots) are achieved at $\alpha^u = 4.5^\circ$ and $\alpha^u = 5^\circ$. The projected arrows show the trajectories of the C points as we increase $\alpha^u$. (c) Calculated Q factors with Q approaching infinity at the $\Gamma$ point for a chiral BIC at $\alpha^u = 5^\circ$. (d) Calculated polarization vortices for a chiral BIC at $\alpha^u = 5^\circ$, where right-handed C points are merging with a Janus BIC in both upward and downward radiation. But for the upward radiation, the LCP C-points do not merge with the BIC at $\Gamma$, whereas the LCP C-points for the downward radiation annihilate with the at-$\Gamma$ BIC, leading to chirality. (e) Calculated $S_3$ for a chiral BIC. Chirality with $|S_3| \geq 0.75$, $|S_3| \geq 0.8$ and $|S_3| \geq 0.95$ is indicated by white contour lines.

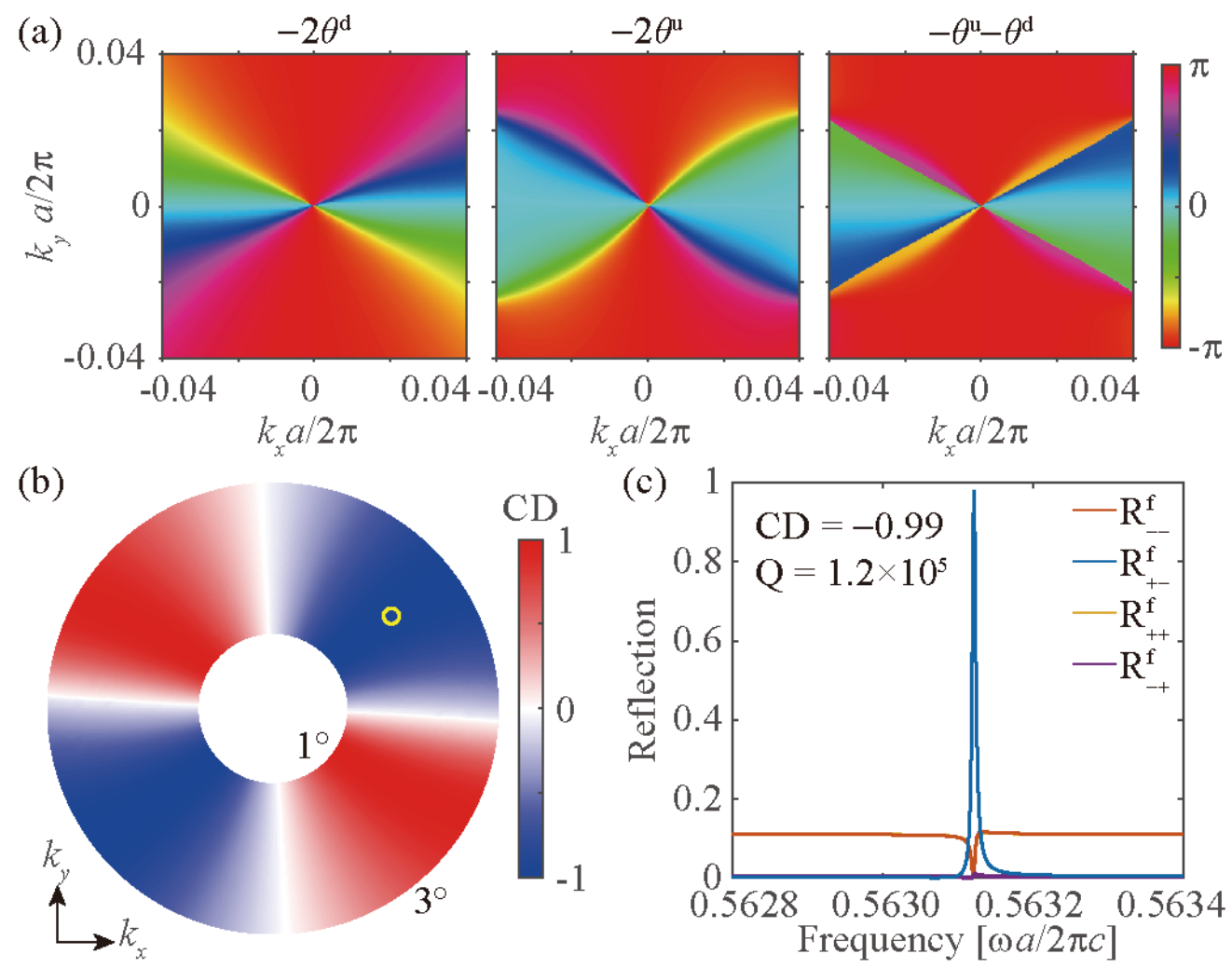

FIG. 4. (a) Calculated Pancharatnam-Berry phase singularities induced by a Janus BIC with $q^{u}=+1$ and $q^{d}=-1$. The parameters used are $\Delta d=12$ nm and $\alpha^{u}=0^{\circ}$. (b) Calculated maximum CD distribution near a chiral BIC with the incident angle from $1^{\circ}$ to $3^{\circ}$ at $\alpha^{u}=5^{\circ}$. (c) Calculated reflection matrix for a chosen incident angle as donated by a yellow circle in (b).